\documentclass[10pt,aps,prb,twocolumn,export,superscriptaddress,nobibnotes,nofootinbib]{revtex4-2}
\pdfoutput=1
\usepackage[utf8]{inputenc}
\usepackage[T1]{fontenc}
\usepackage{anyfontsize}

\usepackage{graphicx}
\usepackage{amsmath}
\usepackage{amssymb}

\usepackage{braket}
\usepackage[dvipsnames,xcdraw]{xcolor}
\usepackage{mathtools} 

\usepackage[export]{adjustbox} % valign
\usepackage[inline]{enumitem}  % inline lists
\usepackage{setspace}

\usepackage{microtype}
\usepackage{newtxtext}
\usepackage[smallerops,subscriptcorrection,varg]{newtxmath}

\DeclareMathAlphabet{\mathcal}{OMS}{cmsy}{m}{n}
\DeclareMathAlphabet{\mathsf}{T1}{cmss}{m}{n}
\DeclareMathAlphabet{\mathbb}{U}{msb}{m}{n}
\usepackage{courierten} % texttt font replace

\usepackage{tikz}
\usetikzlibrary{calc}
\usetikzlibrary{shapes.geometric}
\usetikzlibrary{decorations.markings}
\usetikzlibrary{arrows.meta}

\tikzset{
  SBarb/.tip={Straight Barb[length=2.5pt,width=4.5pt]}, % redefine size
  tikzleftright/.style={<->,>=SBarb}
}

\tikzset{
kb1/.style={postaction={decorate,
   decoration={markings,mark=at position .6 with {\arrow[scale=1.3]{stealth};}}}
   }}

\definecolor{agen}{RGB}{204,41,  41}    % matches eq 18
\definecolor{bgen}{RGB}{41, 106, 204}   % matches eq 18
\definecolor{cgen}{RGB}{100,212, 40}
\definecolor{dgen}{RGB}{229,198, 46}    % matches eq 18

\usepackage{xstring}
\newcommand{\dyckpath}[1]{%
    \begingroup
    \StrLen{#1}[\stringlength]
    \begin{tikzpicture}[x=1ex,y=1ex,baseline={(0, 0)},line cap=round]
        \countdef\dir=255
        \countdef\xpos=256
        \countdef\ypos=257
        \countdef\newx=258
        \countdef\newy=259

        \dir=1
        \xpos=0
        \ypos=0
        \def\lastcolor{}%
        \def\lastnonflatcolor{}%

        \count0=1
        \loop
            \StrChar{#1}{\the\count0}[\thischar]

            \newx=\numexpr\xpos+1\relax
            \newy=\ypos
            \def\thiscolor{black}%

            \IfEq{\thischar}{r}{\def\thiscolor{agen}}{}%
            \IfEq{\thischar}{g}{\def\thiscolor{cgen}}{}%
            \IfEq{\thischar}{b}{\def\thiscolor{bgen}}{}%

            \IfStrEq{\thischar}{e}{}{%
                \ifx\lastnonflatcolor\thischar
                    \multiply\dir by -1
                \fi
                \ifnum\numexpr\ypos+\dir\relax<0
                    \multiply\dir by -1
                \fi
                \newy=\numexpr\ypos+\dir\relax
                \edef\lastnonflatcolor{\thischar}
            }%

            \draw[thick,\thiscolor] (\the\xpos,\the\ypos) -- (\the\newx,\the\newy);

            \xpos=\newx
            \ypos=\newy
            \edef\lastcolor{\thischar}

            \advance\count0 by 1
        \ifnum\count0<\numexpr\stringlength+1\relax
        \repeat
    \end{tikzpicture}%
    \endgroup
}

\usepackage{booktabs} 
\usepackage{tabularx}
\usepackage{nicematrix}
\usepackage[caption=false,subrefformat=parens,labelformat=parens]{subfig}

\usepackage[colorlinks=true,linkcolor=bgen,citecolor=bgen,urlcolor=bgen,pdftitle={Group waffle}]{hyperref}

\usepackage{comment}

\newcommand{\tta}{\mathtt{a}}
\newcommand{\ttb}{\mathtt{b}}
\newcommand{\ttc}{\mathtt{c}}
\newcommand{\ttd}{\mathtt{d}}
\newcommand{\tte}{\mathtt{e}}

\newcommand{\ttg}{\mathtt{g}}
\newcommand{\tth}{\mathtt{h}}

\newcommand{\ttm}{\mathtt{m}}

\newcommand{\tto}{\mathtt{o}}

\newcommand{\ttr}{\mathtt{r}}

\newcommand{\zz}{\mathbb{Z}}

\def\Z#1{\mathbb{Z}_{#1}}

\DeclareMathOperator{\growth}{\mathcal{B}}

\begin{document}

\title{Strong Hilbert space fragmentation and fractons from  subsystem and higher-form symmetries}

\author{\href{https://orcid.org/0000-0002-9809-5575}{\color{black}Charles Stahl}}
\thanks{These two authors contributed equally}
\affiliation{Department of Physics, Stanford University, Stanford, CA 94305, USA}

\author{\href{https://orcid.org/0000-0002-5391-7483}{\color{black}Oliver Hart}}
\thanks{These two authors contributed equally}
\affiliation{Department of Physics and Center for Theory of Quantum Matter, University of Colorado Boulder, Boulder, Colorado 80309 USA}

\author{\href{https://orcid.org/0000-0002-9941-964X}{\color{black}{Alexey Khudorozhkov}}}
\affiliation{Department of Physics, Boston University, Boston, MA 02215, USA}

\author{\href{https://orcid.org/0000-0001-5703-6758}{\color{black}Rahul Nandkishore}}
\affiliation{Department of Physics and Center for Theory of Quantum Matter, University of Colorado Boulder, Boulder, Colorado 80309 USA}

\begin{abstract}
\setstretch{1.1}
We introduce a new route to Hilbert space fragmentation in high dimensions leveraging the group-word formalism. We show that taking strongly fragmented models in one dimension and ``lifting'' to higher dimensions using subsystem symmetries can yield strongly fragmented dynamics in higher dimensions, with subdimensional (e.g., lineonic) excitations. This provides a new route to higher-dimensional strong fragmentation, and also a new route to fractonic behavior. Meanwhile, lifting one-dimensional fragmented models to higher dimensions using higher-form symmetries yields models with topologically robust fragmentation. In three or more spatial dimensions, one can also ``mix and match'' subsystem and higher-form symmetries, leading to canonical fracton models such as X-cube. We speculate that this approach could also yield a new route to non-Abelian fractons. These constructions unify a number of phenomena that have been discussed in the literature, as well as furnishing models with novel properties.
\end{abstract}

\maketitle

\section{Introduction}

Hilbert space fragmentation (HSF, aka Hilbert space shattering) is an exciting new phenomenon~\cite{KHN, Sala2020, Moudgalya2022Thermalization, moudgalya2022quantum}, whereby ergodicity is broken, and the unitary time evolution matrix block diagonalizes into dynamically distinct Krylov sectors, which are not symmetry sectors of any ordinary symmetry. The fragmentation can be ``strong'' (if the largest Krylov sector is a vanishing fraction of the symmetry sector, in the thermodynamic limit), or ``weak'' (if the largest Krylov sector contains almost all of the symmetry sector, in the thermodynamic limit). While HSF is frequently a property of fine-tuned Hamiltonians, it can in certain contexts have a degree of robustness. For instance, in the first-discussed examples~\cite{Sala2020, KHN} HSF is robust to arbitrary spatially local perturbations respecting two symmetries. More recently, ``topological'' HSF was discovered~\cite{stephen2022ergodicity, stahl2023topologically}, wherein the HSF, while weak, is robust to arbitrary $k$-local (but not necessarily spatially local) perturbations. Topological HSF persists at least up to a prethermal timescale exponentially long in a control parameter, and in certain models may even persist to infinite times~\cite{robuster}. 
Very recently, a new group dynamics formalism was introduced~\cite{balasubramanian2023glassy}, which provides a powerful means of treating HSF using ideas from complexity theory. This formalism was employed in Ref.~\cite{AlexeyTopologically} to extend the discussion of topological HSF, originally formulated on square and cubic lattices, to arbitrary crystalline lattices, and even to arbitrary graphs (without translation invariance). These constructions can be understood as a ``lifting'' of one-dimensional HSF to higher dimensions using higher-form symmetries~\cite{Nussinov2007, Gaiotto2015, McGreevy2022}.

In this manuscript we employ the group dynamics formalism of Ref.~\cite{balasubramanian2023glassy} to extend the study of HSF in a new direction, by lifting one-dimensional fragmentation to higher dimensions using \emph{subsystem} symmetries~\cite{VijayHaahFu}. We both provide a general prescription for such subsystem-symmetric liftings, as well as some concrete examples. We show that while lifting one-dimensional models to higher dimensions using higher-form symmetries yields topological HSF, lifting using subsystem symmetries can yield models with strong HSF and fractonic behavior (e.g., involving quasiparticles with restricted mobility). While the previous literature, to our knowledge, contains only two examples of strong HSF in higher dimensions -- the firing/antifiring model of Ref.~\cite{firingantifiring}, recast in the language of multipole symmetries by Ref.~\cite{BHN}, and the ``quantum drums'' model of Ref.~\cite{drums} -- our construction yields access to an infinitely large family of higher-dimensional models exhibiting strong HSF. In three dimensions there is also the possibility of a ``mixed'' lifting involving \emph{both} subsystem and higher-form symmetries, which yields models with topological robustness, but with exponentially more Krylov sectors than pure topological HSF. This type of mixed lifting also yields paradigmatic models of fractons, such as the $\mathbb{Z}_N$ X-cube model, and (we conjecture) could provide a route to the identification of new types of non-Abelian fracton models. Thus, the perspective we introduce enables a unified understanding of a number of models of HSF that exist in the literature, as well as providing new routes to fracton physics, and enabling the discovery of models with new phenomenology. 

\section{Review of group models in 1D}

\subsection{Group dynamics}

Our construction makes use of the recently introduced framework~\cite{balasubramanian2023glassy} for constructing 1D dynamics from a discrete finitely presented group,
\begin{equation}
    G = \langle \mathcal{S} \mkern2mu | \mkern2mu R \rangle .
    \label{eqn:group-presentation}
\end{equation}
For a detailed introduction to the formalism, we refer the reader to Refs.~\cite{balasubramanian2023glassy,AlexeyTopologically}. We consider systems with a finite onsite Hilbert space, whose basis states are labeled by a set of characters $\mathcal{S}$ (e.g., $\mathcal{S} = \{ \tta, \ttb, \ttc \}$), along with their inverses $\mathcal{S}^{-1}$ and an identity character $\tte$. In one dimension (1D), a product state in this basis can be regarded as a \emph{word} constructed from the symmetric alphabet $\mathcal{A}=\mathcal{S} \cup \mathcal{S}^{-1} \cup \{ \tte \}$, formed by concatenating the characters from left to right, e.g., $\ket{\tta\ttb\ttc} \equiv \ket{\tta}\otimes\ket{\ttb}\otimes\ket{\ttc}$ reads the word $w=\tta\ttb\ttc$. 

Viewing the characters as generators of the group $G$ defines a natural, generally noninjective map $\varphi(\,\cdot\,)$ from words to group elements $g \in G$, obtained by multiplying the characters in a word from left to right. Endowing the system with dynamics that preserves the associated group element causes certain states to become dynamically connected, while others remain disconnected. 
This splits the Hilbert space into \emph{Krylov sectors}, each labeled by a distinct group element. 
For Abelian groups, the Krylov sectors are fully labeled by the charges of a conventional Abelian global symmetry~\cite{AlexeyTopologically}. The dynamics constructed this way can be energy-conserving quantum dynamics, random unitary circuits, or even classical stochastic dynamics. 

Simple local dynamics that preserve $\varphi$ can be systematically constructed directly from the presentation~\eqref{eqn:group-presentation}. Free reduction of a character with its inverse and commutation with the identity element are implemented by
\begin{equation}
    {ss^{-1}} \leftrightarrow {\tte\tte} \, ,
    \qquad
    {\tte s} \leftrightarrow {s \tte} \, ,
    \label{eq:free_reduction_dynamics}
\end{equation}
respectively, where $s \in \mathcal{A}$ is an arbitrary character. Nontrivial dynamics descend from the  \emph{relators} $r \in R$ in the group's presentation. For example, the relator $r = \tta \ttb \ttd^{-1} \ttc^{-1}$ means that $\varphi(\tta \ttb \ttd^{-1} \ttc^{-1}) = \tte$, or equivalently $\varphi(\tta \ttb ) = \varphi(\ttc\ttd)$, and therefore product states $\ket{\dots\tta\ttb\dots}$ and $\ket{\dots\ttc\ttd\dots}$, which differ only by this substitution, should be dynamically connected. 
Dynamics corresponding to a generic relator $r \in R$ may be implemented by splitting $r$ into two parts, $r = r^{}_{\text{left}} r^{-1}_{\text{right}}$. This permits replacements of the form ${r_{\text{left}}} \leftrightarrow {r_{\text{right}}}$
in the dynamics, where the shorter of the two words is appropriately padded with $\tte$ characters.

\subsection{Example: 1D Pair-flip model}

The pair-flip model in 1D~\cite{CahaNagaj,moudgalya2022hilbert} can be regarded as deriving from the group $G=\Z{2}^{*m}$ (where $*$ stands for the free product) with presentation
\begin{equation}
    G = \Braket{\tta_1, \tta_2, \dots, \tta_m | \tta_1^2 = \tta_2^2  = \dots = \tta_m^2 = \tte}
    \, ,
\end{equation}
which is different from $\Z{2}^m$ because the generators do not commute. We will consider a pair-flip-inspired group model, including an identity character $\tte$ onsite, unlike the usual pair-flip model in the literature~\cite{CahaNagaj, moudgalya2022hilbert}. See Appendix~\ref{app:with-e} for a comparison to the variant without the $\tte$ character, including the difference in the size of the largest Krylov sector.
The number of group elements that can be realized in a system of length $L$ (the group's \emph{growth function}) scales asymptotically as $\sim \exp(c_1 L)$ with $c_1 = \log(m-1)$~\cite{stahl2023topologically}, which gives rise to an exponential number of Krylov sectors. The largest Krylov sector, which is associated with words satisfying $\varphi(w)=\tte$, grows asymptotically as $\lesssim\exp(c_2 L)$ with $c_2 = \log(1+2\sqrt{m-1})$.
Since the local Hilbert space dimension is $m+1$, this implies that the model exhibits strong fragmentation in the sense of Refs.~\cite{KHN, Sala2020}.

\begin{figure}[t]
    \centering
    \includegraphics[width=\linewidth]{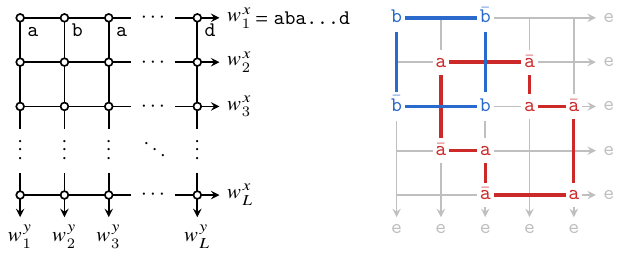}
    \caption{(Left) The orientation conventions used to define 2D subsystem group models. The dynamics preserve the group elements $\varphi(w^\alpha_{i})$ associated with the words $w^{\alpha}_{i}$ read along each of the rows ($\alpha=x$, read from left to right) and columns ($\alpha=y$, read from top to bottom) of a 2D square lattice. Row and column indices satisfy $1 \leq i \leq L$ and open boundary conditions are imposed throughout the paper. (Right) A configuration belonging to the sector in which every row and column satisfies $\varphi(w_i^\alpha) = \tte$. A bar above a character denotes the inverse.}
    \label{fig:2D-subsys-conventions}
\end{figure}

\section{Subsystem group models in 2D}

\subsection{Subsystem group dynamics}

We now lift one-dimensional group models to two dimensions using subsystem symmetries.
Consider a two-dimensional square lattice with degrees of freedom on the vertices.
Furthermore, every row and column is endowed with an orientation, which specifies the order in which characters are concatenated to form a word. In 1D, this orientation was implicitly chosen such that characters were concatenated from left to right. In 2D, we work with the convention shown in Fig.~\ref{fig:2D-subsys-conventions}, which defines a set of words $w^x_{i}$ and $w^y_i$, with $i=1, \dots, L$, reading along the rows and columns, respectively.\footnote{For certain groups, it is possible that different conventions for the orientation of rows and columns will result in different allowed dynamics. We leave this possibility to future work.}
We then consider dynamics that simultaneously preserves the group elements $\varphi(w^{\alpha
}_i)$ for all $i$, $\alpha$, i.e., for all rows and columns. 
We will refer to such dynamics as \emph{$\varphi$-preserving}, and the set of group elements on rows and columns as the \emph{subsystem group label}. 

Simple $\varphi$-preserving dynamics analogous to free reduction and commutation with the identity~\eqref{eq:free_reduction_dynamics} can be implemented as
\begin{equation}
    \begin{bmatrix} \, s^{\phantom{-1}} & s^{-1} \\ \, s^{-1} & s^{\phantom{-1}} \end{bmatrix} \leftrightarrow
    \begin{bmatrix} \tte & \tte \, \\ \tte & \tte\, \end{bmatrix}
    , 
    \qquad
    \begin{bmatrix} \, s & \tte \\ \, \tte & s \end{bmatrix} \leftrightarrow
    \begin{bmatrix} \tte & s \, \\ s & \tte\, \end{bmatrix}
    ,
    \label{eq:free_reduction_2D}
\end{equation}
respectively, where $s \in \mathcal{A}$. Here, the first operator nucleates a quadrupole of generator $s$, while the second one performs a ``ring-exchange'' update. Related operators can either give dynamics to a dipole of the same generator or expand a monopole into three characters, if the neighboring sites are unoccupied:
\begin{equation}
    \begin{bmatrix} \, \tte & s^{\phantom{-1}} \\ \, \tte & s^{-1} \end{bmatrix} \leftrightarrow
    \begin{bmatrix} s^{\phantom{-1}} & \tte \, \\ s^{-1} & \tte \, \end{bmatrix}\,
    ,
    \qquad
    \begin{bmatrix} \, s^{\phantom{-1}} & \tte \\ \, s^{-1} & s \end{bmatrix}\leftrightarrow
    \begin{bmatrix} \tte & s \, \\ \tte & \tte \,\end{bmatrix}\,
    .
    \label{eq:dipole_movement_2D}
\end{equation}
Generic local dynamics deriving from free reduction are related to these operators by rotation and reflection.
Acting repeatedly with these operators can spread a quadrupole out to the corners of a rectangle, or create a sequence of alternating $s$ and $s^{-1}$ in the corners of an arbitrary rectilinear region (see Fig.~\ref{fig:2D-subsys-conventions}).
More generally, any word with $\varphi(w) = \tte$, if embedded in a local vacuum, acts as a lineon -- an excitation that can freely move through the vacuum in one direction (along the line perpendicular to the word), but not in the other direction (parallel to the word). Such ``lineonic'' behavior is a hallmark of fracton models \cite{castelnovo2010quantum, castelnovo2012topological, vijay2015new, VijayHaahFu, Haah, nandkishore2019fractons}.  At the same time, any local patch of words that has $\varphi(w) = \tte$ on every row and column [as in Eq.~\eqref{eqn:abc_grid_flip}] acts as a fully mobile particle and can be locally created or destroyed.

If our presentation includes nontrivial relations, then these relations lead to other classes of local dynamics that leave the subsystem group label invariant.
For example, the relation $\tta \ttb \ttc = \tte$ would generate dynamics such as
\begin{equation}
\begin{bmatrix} \tte & \tte  & \tte \\ \tte & \tte & \tte \\ \tte & \tte & \tte\end{bmatrix}  \longleftrightarrow  
\begin{bmatrix} \tta & \ttb^{\phantom{-1}} & \ttc^{\phantom{-1}} \\ \ttb & \ttb^{-1} &  \\ \ttc & & \ttc^{-1} \end{bmatrix}
 \longleftrightarrow  
\begin{bmatrix} \tta^{-1} & & \tta \\ & \ttb^{-1} & \ttb \\ \tta^{\phantom{-1}} & \ttb^{\phantom{-1}} & \ttc \end{bmatrix}
\, ,
\label{eqn:abc_grid_flip}
\end{equation}
where blank spaces imply $\tte$ characters. If there is a relation of the form $\tta\ttr\ttg\tth = \ttm\ttg\tto\ttg$, $\varphi$-preserving dynamics include
\begin{equation}
\begin{bmatrix}
\tta^{-1} & & \tta & \\ & \ttr^{-1} & \ttr & \\ \tta^{\phantom{-1}} & \ttr^{\phantom{-1}} & \ttg & \tth^{\phantom{-1}} \\ & & \tth & \tth^{-1}
\end{bmatrix} 
\longleftrightarrow 
\begin{bmatrix}
 \ttm^{-1} & & \ttm & \\ & \ttg^{-1} & \ttg & \\ \ttm^{\phantom{-1}} & \ttg^{\phantom{-1}} & \tto & \ttg^{\phantom{-1}} \\ & & \ttg & \ttg^{-1}
\end{bmatrix} 
\end{equation}
where a nontrivial word is read along the third row and column. The prescription for constructing similar dynamical terms proceeds as follows: start with a word $w$, and place two copies of it -- one horizontally and one vertically -- so that they intersect on a shared character.\footnote{If the word contains repeating characters, then the two copies can intersect on characters in different positions.} Then, for each remaining pair of equal characters, insert the inverse of that character at the intersection of the corresponding row and column. Such a pattern can be flipped to an analogous one built from a word $\tilde{w}$ with $\varphi(\tilde{w}) = \varphi(w)$, keeping the intersection point fixed at the same vertex.

If the group -- and its specific presentation -- has additional structure, it may allow for dynamical terms beyond the pattern described above. For instance, any \emph{palindromic relation}, meaning a relation of the form $\varphi(\tta_1 \tta_2 \dots \tta_n) = \varphi(\ttb_1 \ttb_2 \dots \ttb_n)$ such that $\varphi(\tta_n \dots \tta_2 \tta_1) = \varphi(\ttb_n \dots \ttb_2 \ttb_1)$ (i.e., with the order reversed), gives rise to dynamics of the form
\begin{equation}
\begin{bmatrix} 
    \tta_1^{\phantom{-1}} & \tta_2^{\phantom{-1}} & \cdots & \tta_n^{\phantom{-1}} \\
    \tta_1^{-1}           & \tta_2^{-1}           & \cdots & \tta_n^{-1} 
\end{bmatrix}  
\longleftrightarrow  
\begin{bmatrix}
    \ttb_1^{\phantom{-1}} & \ttb_2^{\phantom{-1}} & \cdots & \ttb_n^{\phantom{-1}} \\
    \ttb_1^{-1}           & \ttb_2^{-1}           & \cdots & \ttb_n^{-1} 
\end{bmatrix}
\, ,
\label{eqn:horiz_palindrome}
\end{equation}
and analogous replacements with words oriented along columns.
Such dynamics are generally not allowed for arbitrary presentations of arbitrary groups.\footnote{It would however be allowed for arbitrary groups given a \textit{boustrophedon} orientation convention, where even and odd rows are read from left to right and from right to left, respectively.} 

Under $\varphi$-preserving dynamics, Hilbert space fragments into disconnected Krylov sectors. \emph{Intrinsic} Krylov sectors are labeled by the subsystem group label.
For a group with growth rate $\growth(L)$, the number of intrinsic Krylov sectors belonging to a system of size $L$ by $W$ is bounded from below by 
\begin{equation}
\max \left(\growth(L)^W, \growth(W)^L \right),
\end{equation}
by independently choosing the group element on each row or on each column.
These models can also exhibit \emph{fragile HSF} as introduced in Refs.~\cite{balasubramanian2023glassy,AlexeyTopologically}, a jamming-type phenomenon that occurs if any row or column possesses a word whose expansion length is longer than the system size. In 1D, specifying all the relations fully specifies the dynamics. 
We will see that this is not the case in the subsystem group models, so that the structure of the HSF depends on both the choice of group and the gate range.

\subsection{Strong fragmentation in subsystem group models}

At this point, we observe that if 1D group dynamics is strongly fragmented for a group $G$, then subsystem group dynamics for $G$ automatically generates strong HSF in higher dimensions. This may be shown by induction. Start with strongly fragmented dynamics in 1D. By postulate, the size of the largest Krylov sector $\dim(K^1_{\mathrm{max}})$ is a vanishing fraction of the Hilbert space dimension $\dim(H^1)$ in the thermodynamic limit, where the superscript refers to the dimensionality of the system. Now lift to 2D, enforcing the constraints of the 1D dynamics along every row. This produces a model of decoupled 1D chains. If there are $L$ rows, then the total Hilbert space has size $\dim(H^2) = \dim(H^1)^L$, while the largest Krylov sector has dimension $\dim(K^2_{\mathrm{max}}) = \dim(K^1_{\mathrm{max}})^L$ . It immediately follows that $\dim(K^2_{\mathrm{max}})/\dim(H^2) = [\dim(K^1_{\mathrm{max}})/\dim(H^1) ]^L \rightarrow 0$. Now, turn on constraints equivalent to those of the 1D dynamics along every column. This further constrains the dynamics, which cannot increase the dimension of the largest Krylov sector. Consequently, 2D dynamics with the constraints of the parent 1D dynamics applied along every row and column must have $\dim(K^2_{\mathrm{max}})/\dim(H^2) \rightarrow 0$, indicating strong fragmentation.\footnote{Strictly, we should compare $\dim(K_{\mathrm{max}})$ to the dimension of the symmetry sector if any global symmetries are present, but symmetry sector size will generically only differ from $\dim(H)$ by a polynomial factor, so the exponential separation demonstrated above will continue to hold.}

This construction may be inductively extended to any number of dimensions $\ge 2$ and it is immediately obvious that if the constraints of the parent strongly fragmented dynamics are applied along every linelike subsystem, then the resulting higher-dimensional dynamics must also exhibit strong HSF. Thus, while previously only two examples of strong HSF in higher dimensions were known \cite{firingantifiring, drums}, our subsystem symmetric lifting prescription provides a route to many more higher-dimensional models with strong HSF, with the subsystem-symmetric lifting of the pair-flip group model outlined below providing a concrete example.

\subsection{Example: 2D subsystem pair-flip model}

As an example, consider the group $G = \mathbb{Z}_2^{*3}$, which gives a 2D subsystem version of the pair-flip model.
The onsite Hilbert space is spanned by the alphabet $\mathcal{A} = \{ \tte, {\color{agen}\ttr}, {\color{bgen}\ttg}, {\color{cgen}\ttb} \}$. As in 1D, we choose to work with an $\tte$ character in the local Hilbert space because of the simpler dynamics, but the results for the subsystem pair-flip model without an onsite $\tte$ character should be similar. For example, as discussed in Appendix~\ref{app:with-e}, the Krylov sector labeling structure of the two models is identical.
Examples of allowed dynamics include the terms derived from free reduction, 
\begin{equation}
\begin{bmatrix}
{\tte} & {\tte} \\ {\tte} & {\tte} 
\end{bmatrix} 
\longleftrightarrow 
\begin{bmatrix}
{\color{agen}\ttr} & {\color{agen}\ttr} \\ {\color{agen}\ttr} & {\color{agen}\ttr} \\
\end{bmatrix} 
\longleftrightarrow 
\begin{bmatrix}
{\color{bgen}\ttb} & {\color{bgen}\ttb} \\ {\color{bgen}\ttb} & {\color{bgen}\ttb} \\
\end{bmatrix}
\longleftrightarrow 
\begin{bmatrix}
{\color{cgen}\ttg} & {\color{cgen}\ttg} \\ {\color{cgen}\ttg} & {\color{cgen}\ttg} \\
\end{bmatrix}, 
\label{eq:pairflip_free_reduction}
\end{equation}
commutation with the identity,
\begin{equation}
\begin{bmatrix}
{\tte} & {\tte} \\ {\color{agen}\ttr} & {\tte} 
\end{bmatrix} 
\longleftrightarrow 
\begin{bmatrix}
{\color{agen}\ttr} & {\color{agen}\ttr} \\ {\tte} & {\color{agen}\ttr} \\
\end{bmatrix} , \qquad
\begin{bmatrix}
{\tte} & {\color{agen}\ttr} \\ {\color{agen}\ttr} & {\tte} 
\end{bmatrix} 
\longleftrightarrow 
\begin{bmatrix}
{\color{agen}\ttr} & {\tte} \\ {\tte} & {\color{agen}\ttr} \\
\end{bmatrix},
\label{eq:pairflip_free_reduction_2}
\end{equation}
as well as relations related by spacial rotation or by changing $\color{agen}\ttr$ to $\color{cgen}\ttg$ or $\color{bgen}\ttb$. These dynamics can be composed to realize more complicated terms, such as,
\begin{equation}
\begin{bmatrix}
{\color{agen}\ttr} & {\color{bgen}\ttb} & {\color{cgen}\ttg} & {\color{cgen}\ttg} \\ {\color{bgen}\ttb} & {\color{cgen}\ttg} & {\color{cgen}\ttg} & {\color{agen}\ttr} \\ {\color{bgen}\ttb} & {\color{cgen}\ttg} & {\color{agen}\ttr} & {\color{agen}\ttr} \\ {\color{agen}\ttr} & {\color{bgen}\ttb} & {\color{agen}\ttr} & {\color{agen}\ttr}
\end{bmatrix} 
\longleftrightarrow 
\begin{bmatrix}
{\color{agen}\ttr} & {\color{bgen}\ttb} & {\color{cgen}\ttg} & {\color{cgen}\ttg} \\ {\color{bgen}\ttb} & {\color{agen}\ttr} & {\color{agen}\ttr} & {\color{agen}\ttr} \\ {\color{bgen}\ttb} & {\color{agen}\ttr} & {\color{agen}\ttr} & {\color{cgen}\ttg} \\ {\color{agen}\ttr} & {\color{bgen}\ttb} & {\color{cgen}\ttg} & {\color{cgen}\ttg}
\end{bmatrix} 
\longleftrightarrow 
\begin{bmatrix}
{\color{agen}\ttr} & {\color{bgen}\ttb} & {\color{cgen}\ttg} & {\color{cgen}\ttg} \\ {\color{agen}\ttr} & {\color{agen}\ttr} & {\color{bgen}\ttb} & {\color{agen}\ttr} \\ {\color{agen}\ttr} & {\color{agen}\ttr} & {\color{bgen}\ttb} & {\color{cgen}\ttg} \\ {\color{agen}\ttr} & {\color{bgen}\ttb} & {\color{cgen}\ttg} & {\color{cgen}\ttg}
\end{bmatrix} , 
\label{eq:pairflip_complicated_relation}
\end{equation}
where, in each case, the horizontal group elements are $\color{agen}\ttr \color{bgen}\ttb$, $\color{bgen}\ttb \color{agen}\ttr$, $\color{bgen}\ttb \color{cgen}\ttg$, and $\color{agen}\ttr \color{bgen}\ttb$, while the vertical elements are $\tte$, $\tte$, $\tte$, and $\color{cgen}\ttg \color{agen}\ttr$. Since, as we outlined above, the 1D pair-flip model is strongly fragmented, its lifting to two dimensions using subsystem symmetries results in a strongly fragmented 2D model with the size of the largest Krylov sector scaling as 
\begin{equation}
    \frac{\dim(K_{\mathrm{max}})}{\dim(H)} \lesssim \left( \frac{1+2\sqrt{m-1}}{m+1} \right)^{L^2} \to 0 \quad \text{at}\enspace L \to \infty,
\end{equation}
for all $m \geq 3$, where the example above was written for $m=3$.

Unlike in 1D, the minimal dynamics coming from the relations do not generate all local $\varphi$-preserving dynamics. For example, the two configurations
\begin{equation}
\begin{bmatrix}
\tte & \tte & \tte & \tte \\
\tte & \tte & \tte & \tte \\
\tte & \tte & \tte & \tte \\
\tte & \tte & \tte & \tte 
\end{bmatrix} 
\xleftrightarrow{\quad ?\quad}
\begin{bmatrix}
{\color{agen}\ttr} & \tte & {\color{agen}\ttr} & \tte \\ \tte & {\color{bgen}\ttb} & \tte & {\color{bgen}\ttb} \\ {\color{agen}\ttr} & \tte & {\color{agen}\ttr} & \tte \\ \tte & {\color{bgen}\ttb} & \tte & {\color{bgen}\ttb}
\end{bmatrix} , 
\label{eq:pairflip_linked_loops}
\end{equation}
have the same (trivial) subsystem group label, but cannot be connected by range-2 $\varphi$-preserving gates. Thus, to fully specify what we mean by ``subsystem group dynamics'' we have to specify a group $G$ and either an explicit set of $\varphi$-preserving gates or a gate range. In other words, the subsystem group label does not in general fully label the Krylov sector structure.  Among states where the subsystem group label is trivial, the complete labeling of Krylov sectors reduces to finding equivalence classes of a 2D Dyck language, a problem that has received attention in recent math literature~\cite{subsystemwordproblems}. For a generic group and without restriction to sectors with trivial subsystem group label, it reduces to identifying equivalence classes of a 2D ``picture language,'' which is a formidably difficult (even undecidable) task~\cite{2dwordproblems}.

\subsection{Subsystem semigroup models}

The construction can be conceptually extended from groups to semigroups by considering a fixed semigroup presentation and requiring that the dynamics preserve the semigroup element along each row and column. In practice, however, writing down generic dynamical terms for an arbitrary semigroup is challenging due to the lack of an identity and/or inverse elements.

Nevertheless, in specific cases, one can perform such lifting, and if the corresponding 1D model is strongly fragmented, then the resulting 2D model will be as well. Below we introduce a new two-dimensional strongly fragmented model, based on a subsystem symmetric lifting of the paradigmatic charge-and-dipole conserving model \cite{KHN, Sala2020}. 
Consider a system on a square lattice with spin-$1/2$ variables placed at every vertex. Now suppose that the dynamics is locally generated and conserves both charge (total $S^z$) and the dipole moment of charge along every row and column. Such dynamics may be generated by operations such as 
\begin{equation}
\renewcommand{\arraystretch}{0.8}
\setlength{\arraycolsep}{1.5pt}
\begin{bmatrix}
+ & - & - & + \\ - & + & + & - \\ - & + & + & - \\ + & - & - & +
\end{bmatrix} 
\longleftrightarrow 
\begin{bmatrix}
- & + & + & - \\+ & - & - & +  \\ + & - & - & + \\ - & + & + & -
\end{bmatrix} 
\end{equation}
where $+$ ($-$) denotes a site with $S^z = + 1/2$ ($S^z = -1/2$). The ``parent'' dynamics being applied along any row or column is the elementary charge and  dipole conserving dynamics for a spin-$1/2$ system with four-site gates, which is strongly fragmented~\cite{Sala2020}. The lifting thereof generates a novel strongly fragmented dynamics in two dimensions. The construction may be further lifted to three (or more) dimensions by taking the outer product of the dynamics above with the elementary 1D dynamics ($+$$-$$-$$+$ $\leftrightarrow$ $-$$+$$+$$-$). Hence, one may generate strongly fragmented dynamics in arbitrary dimensions via a subsystem-symmetric lifting of the charge-dipole problem. In fact, this construction produces dynamics that is not only strongly fragmented, but is \emph{localized}. The parent model in one dimension has finite-thickness ``bottleneck'' motifs~\cite{KHN}, and any $d$-dimensional region fully surrounded by the bottleneck motifs will be completely disconnected from the rest of the system, with all charges internal to said region remaining in that region for all time. Since any finite-sized region has nonzero probability of being surrounded by bottleneck motifs, we should expect that typical initial conditions will manifest a memory of the initial conditions in local observables for all times, yielding (statistical) localization~\cite{slioms}.

\section{Robustness from higher-form subsystem symmetries}

Above we have discussed liftings employing subsystem symmetry and producing strong fragmentation. These constructions are not robust to generic (subsystem asymmetric) perturbations. In previous literature~\cite{stahl2023topologically, AlexeyTopologically} an alternative lifting based on higher-form symmetries has been discussed, which produces topologically robust weak fragmentation. The robust lifted models live on 2D lattices with degrees of freedom on edges, and the robustness comes from requiring that the words read on contractible paths on the lattice map to the identity element. We call this condition \emph{flatness}, because restricting this construction to finite groups exactly reproduces the flatness condition of discrete gauge theory. Like in gauge theory, the flatness constraints lead to the emergence of higher-form symmetries characterized by deformable subdimensional symmetry operators~\cite{Nussinov2007, Gaiotto2015, McGreevy2022}. For further details, see Ref.~\cite{AlexeyTopologically}.

\begin{figure}[t]
    \centering
    \includegraphics[valign=c]{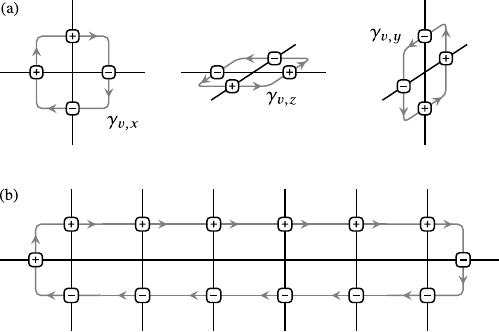}
    \caption{(a) In the 3D subsystem symmetry model, impose flatness on the paths $\gamma_{v, x}$, $\gamma_{v, y}$, and $\gamma_{v, z}$ around every vertex $v$. Proceeding in the direction indicated by the gray arrows, read off the character on edges with a $+$ and the inverse character on edges with a $-$. In flat configurations the resulting word maps to the identity group element. (b) When combined, the constraints require that the words read along two parallel rows map to the same group element.}
    \label{fig:fractonflat}
\end{figure}

We now consider mixed liftings involving both subsystem and higher-form symmetries. This requires working in at least three spatial dimensions. Consider a system with a finite Hilbert space associated to each edge of a translationally invariant lattice. For concreteness, consider the cubic lattice. As before, let the onsite Hilbert space consist of states $\ket{s}$, for $s\in \mathcal{A}$. Now impose flatness on every X-shaped configuration of edges around a vertex, defined as in Fig. \ref{fig:fractonflat}.
This particular choice is inspired by the $\zz_N$ X-cube model in Ref.~\cite{VijayIsotropic}. 
The operators that detect violations of flatness around vertex $v$ are 
\begin{equation} \label{eqn:flatness-operator}
    B_{v,\alpha} = \sum_{ h_{e_1} h_{e_2} h_{e_3} h_{e_4} =\tte} \: \prod_{e \in \gamma_{v, \alpha}} T^{h_e}_{v,\alpha}(e)
    \, , 
\end{equation}
where $\alpha = x, y, z$ labels one of the principal lattice directions and $\gamma$ is one of the paths defined in Fig.~\ref{fig:fractonflat}(a). 
The operator $T^{h_e}_{v,\alpha}(e)$ is either $|h\rangle \langle h|$ or $| h^{-1} \rangle \langle h^{-1} |$, as defined in Fig.~\ref{fig:fractonflat}.
Effectively, $B_{v, \alpha}=1$ requires that the word read from the four edges around the vertex $v$ in the direction specified by $\gamma$ maps to the identity. We call $B_{v,\alpha}=1$ a higher-form constraint because it reduces to a higher-form symmetry when $G$ is Abelian~\cite{AlexeyTopologically}.

A horizontal 2D slice through a layer of the system reveals a square grid of vertical edges, which can be interpreted as a 2D subsystem group model. We can read group elements off of rigid 1D slices of this system [see Fig.~\ref{fig:fractonflat}(b)] which correspond to 2D subsystem group labels. Naively, there are $O(L)$ such labels per layer, but
the flatness constraints ensure that the words read along parallel paths in different layers correspond to the same group element, so they are not all independent. Instead, there are only $O(L)$ group labels total. Furthermore, the fact that labels on different layers must match gives rise to topological invariance of the group labels;
The only way to change the label on any row without violating constraints is to change all $L$ parallel rows simultaneously. Such an operator has weight at least $L$. The flatness conditions impose constraints on the system so that any dynamics that preserve flatness also preserve group labels. Thus, in these models, the subsystem group labels are emergent.

The minimal constraint-obeying dynamics acts on the 12 edges around a cube, 
\begin{equation}
A_c(\tta): \; \vcenter{\hbox{\includegraphics[scale=.7]{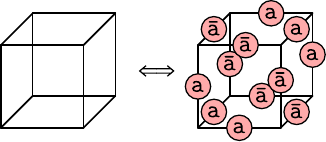}}}, \label{eq:3d-free}
\end{equation}
as in the construction of Ref.~\cite{VijayIsotropic}. Note that, restricting to any four parallel edges of the cube, this dynamics reproduces~\eqref{eq:free_reduction_2D}. As a consequence, it does not change any of the Krylov labels. 

This interpretation suggests a way to construct new dynamics: Take a dynamical term from the 2D subsystem construction, apply the gate to vertical edges in a single layer, and ``close up'' the dangling vertices so as to satisfy all of the constraints. For example, taking the gate in~\eqref{eqn:abc_grid_flip}, we can construct the 3D configuration
\begin{equation}
\vcenter{\hbox{\includegraphics[scale=.5]{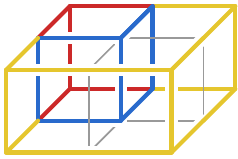}}}\,,
\end{equation}
where we have used red edges to represent $\tta$ and $\tta^{-1}$, blue edges to represent $\ttb$ and $\ttb^{-1}$, and yellow edges to represent $\ttc$ and $\ttc^{-1}$.
Edges where multiple colors meet rely on the relation $\tta\ttb\ttc=\tte$. When there are no relations between different characters, as in the subsystem pair-flip model above, it is still possible to ``close up'' vertices by going to higher layers. As an example, the configuration 
\begin{equation}
\vcenter{\hbox{\includegraphics[scale=.5]{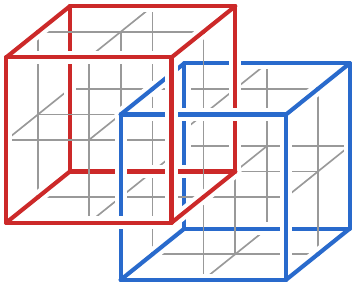}}}
\end{equation}
descends from the dynamics in~\eqref{eq:pairflip_linked_loops}.
Generic constraint-satisfying dynamics can be constructed from all of the allowed gates in 2D, and these operators will be arbitrarily large. Just as the operator in~\eqref{eq:3d-free} creates a cage shape, generic dynamics will create cage-nets, as in Ref.~\cite{Prem2019}.

As in Refs.~\cite{AlexeyTopologically, robuster}, we can construct quantum Hamiltonians by including the $B_{v, \alpha}$ as energetic constraints. Then, generic low-energy effective Hamiltonians implement the constraint-obeying dynamics.
While this generically results in a complicated construction due to large irreducible gates, we gain simplicity by considering finite Abelian $G$ and the presentation $\mathcal{A}_G$ with a character for every group element. Then, the cage operators are completely generic, and the Hamiltonian
\begin{align}
    H_{\text{X-cube}}(G) = -\Delta \sum_{v, \alpha} B_{v,\alpha} - \lambda \sum_{c,g \in G} A_c(g)
    \label{eqn:X-cube-generic}
\end{align}
is exactly solvable. 
When $G=\zz_N$, this Hamiltonian is the $\zz_N$ X-cube model, whose ground states realize fracton topological order~\cite{VijayHaahFu, VijayIsotropic, Slagle2018}. This example demonstrates that, while our models generically exhibit fractonic dynamics, they also reduce to models with fractonic ground states in certain cases. Whether this construction can lead to non-Abelian fractons~\cite{Vijay2017NonAbelian, Prem2019, Bulmash2019, Aasen2020, Tantivasadakarn2021, Williamson2023, Tu2021}, or other novel types of fracton order, remains an interesting open question.

\begin{table}[t]
\def\arraystretch{1.4}
\caption{Overview of the models considered in this manuscript. Spatial dimension is denoted by $d$. The parameter $p$ controls the strength of shattering and its robustness. The number of Krylov sectors scales as $\exp[O(L^{p})]$, and strong fragmentation can occur for $d=p$. Operators that change Krylov sector act on $O(L^{d-p})$ degrees of freedom such that, for $p < d$, the fragmentation is robust.}
\begin{NiceTabularX}{\columnwidth}{X[l]X[l]X[l]X[l]X[l]}
\toprule
      & $p=0$    & $p=1$   & $p=2$  & $p=3$    \\ \midrule
$d=1$ & 1D Ising & Ref.~\cite{balasubramanian2023glassy} &           &          \\
$d=2$ & 2D Ising & Ref.~\cite{AlexeyTopologically}       & Subsystem &           \\
$d=3$ & 3D Ising & Ref.~\cite{AlexeyTopologically}       & Mixed     &  Subsystem \\ \bottomrule
\CodeAfter
  \tikz \draw (2-|4) -- (4-|6) ; 
\end{NiceTabularX}
\label{tab:models}
\end{table}

This section completes a collection of models parametrized by the spatial dimension $d$ and the codimension of Krylov-sector-changing operators $p$. For $1 \le d \le 3$ and $0 \le p \le d$, our constructions give rise to models with $O(L^p)$ Krylov sectors such that operators that change the Krylov sector must act on $O(L^{d-p})$ sites, summarized in Table~\ref{tab:models}. For each of these models, if a global symmetry exists, it is a $p$-form symmetry. Thus, $p=0$ produces models with ordinary global symmetries.  The models with $p = 1$ and $d>p$ are the topologically fragmented models of Refs.~\cite{stahl2023topologically,AlexeyTopologically}. Meanwhile, $d=p$ can yield strong fragmentation, including in higher dimensions, as introduced herein. Finally, $p=2, d=3$ constitutes the mixed lifting (i.e., employing \emph{both} higher-form and subsystem symmetries) that, e.g., produces the $\mathbb{Z}_N$ X-cube model~\eqref{eqn:X-cube-generic}. Such mixed liftings have topological robustness, but also have more Krylov sectors than one would obtain in pure topological fragmentation.

\section{Outlook}

We have introduced a new ``lifting'' of one-dimensional fragmentation to higher dimensions using subsystem symmetry, potentially combined with higher-form symmetry. We have shown that lifting strongly fragmented dynamics to higher dimensions using purely subsystem symmetries produces strongly fragmented dynamics with lineon excitations in higher dimensions. Meanwhile, liftings that combine subsystem and higher-form symmetries produce models with greater robustness, and also lead to paradigmatic fracton models. The constructions presented herein unify a number of phenomena that have been separately discussed in the literature, as well as providing new routes to exotic phenomenology such as strong HSF in higher dimensions, and fractonic behavior. 

A number of directions present themselves for future study. For instance, we have considered here only subsystems with integer dimension. A generalization to fractal subsystems, as in the Newman-Moore model \cite{NewmanMoore}, Castelnovo-Chamon model on the hexagonal close-packed lattice \cite{castelnovo2012topological}, or Haah's code \cite{Haah} would be an interesting topic for future work, and may well yield novel phenomenology. It would also be interesting to extend these constructions to non-Abelian symmetries. This would require a consistent way to define non-Abelian subsystem symmetries in higher dimensions, but may well offer a new route to, e.g., non-Abelian fractons. Finally, while we have focused herein on Krylov sector structure, the full exploration of a given Krylov sector can itself take a very long time \cite{balasubramanian2023glassy}, and exploration of such timescales in the present context could itself be interesting. 

\begin{acknowledgments}
R.N.\ and O.H.\ are supported by the U.S.\ Department of Energy, Office of Science, Basic Energy Sciences under Award No.\ DE-SC0021346. C.S.\ is supported by Vedika Khemani through the Office of Naval Research Young Investigator Program (ONR YIP) under Award Number N00014-24-1-2098. AK is supported by the National Science Foundation under the grant GCR-2428487.
\end{acknowledgments}

\appendix

\section{Modifying the onsite Hilbert space}
\label{app:with-e}

\subsection{1D pair-flip model}

Here we review the dynamics of the 1D pair-flip group model, which includes an identity character ($\tte$) onsite, and describe how the enumeration of states belonging to a specific Krylov sector is modified compared to the conventional pair-flip model. The onsite Hilbert space of the pair-flip group model is spanned by $\mathcal{A} = \{ \tte, {\color{agen}\ttr}, {\color{bgen}\ttb}, {\color{cgen}\ttg}, \dots \}$, where the dots denote the possible inclusion of more colors. The dynamics of the pair-flip group model (with $\tte$, black arrows) and the conventional pair-flip model (without $\tte$, gray arrows) can be summarized by: 
\begin{equation}
\begin{tikzpicture}[x=6ex, y=6ex, baseline={(0, 0)}]
    \node (E) at (0,0) {$\tte \tte$};
    \node (R) at (0:1) {$\color{agen}\ttr \ttr$};
    \node (B) at (120:1) {$\color{bgen}\ttb \ttb$};
    \node (G) at (240:1) {$\color{cgen}\ttg \ttg$};
    \draw [thick, tikzleftright] (E)--(R);
    \draw [thick, tikzleftright] (E)--(B);
    \draw [thick, tikzleftright] (E)--(G);
    \draw[thick, lightgray, tikzleftright] (15:1) arc[start angle=15, end angle=105, radius=1];
    \draw[thick, lightgray, tikzleftright] (135:1) arc[start angle=135, end angle=225, radius=1];
    \draw[thick, lightgray, tikzleftright] (255:1) arc[start angle=255, end angle=345, radius=1];
    \node at (3, 0.66) {\strut${{\color{agen}\ttr} \tte} \leftrightarrow {\tte {\color{agen}\ttr}}$};
    \node at (3, 0.0) {\strut${{\color{bgen}\ttb} \tte} \leftrightarrow {\tte {\color{bgen}\ttb}}$};
    \node at (3, -0.66) {\strut${{\color{cgen}\ttg} \tte} \leftrightarrow {\tte {\color{cgen}\ttg}}$};
\end{tikzpicture}
\label{eqn:1D-PF-dynamics}
\end{equation}
Transitions ${\color{agen}\ttr \ttr } \leftrightarrow {\color{bgen}\ttb \ttb} \leftrightarrow {\color{cgen}\ttg \ttg}$ are possible in both models, but are facilitated by the $\tte\tte$ configuration in the group-model case. The main difference between the dynamics of the two models therefore stems from commutation with the identity character, illustrated on the right-hand side of Eq.~\eqref{eqn:1D-PF-dynamics}. An isolated character, say ${\color{agen}\ttr}$, in a background of $\tte$ characters can transition to any site. However, in the conventional pair-flip model, an isolated ${\color{agen}\ttr}$ in a background of (say) ${\color{bgen}\ttb}$ characters can only move to sites belonging to the same sublattice, i.e., via ${\color{agen}\ttr}{\color{bgen}\ttb\ttb} \leftrightarrow {\color{agen}\ttr\ttr\ttr} \leftrightarrow {\color{bgen}\ttb\ttb}{\color{agen}\ttr}$.

% \begin{equation}
%     \dyckpath{rereggrgbergbeebeegrebegree}
% \end{equation}

For the conventional PF model, evaluating the size of largest sector is equivalent to enumerating colored Dyck paths~\cite{CahaNagaj}, where the number of steps in the path is equal to the length of the 1D system. Including an identity character leads to an analogous enumeration problem in which $\tte$ characters are associated to horizontal steps. One can then show that the dimension of the Krylov sector associated with the identity element is enumerated by the generating function 
\begin{align}
    G_{e}(x) &= \frac{2 (m-1)}{(m-2)(1-x)+m \sqrt{1-2x+(5-4 m) x^2}} \\
    &= 1 + x + (m+1)x^2 + (3m+1)x^3 + \dots \nonumber \\
    &\sim \left\{ 
        \dyckpath{e}
    , m
        \dyckpath{rr}\!+\!\dyckpath{ee}
    , m(
        \dyckpath{err}\!+\!\dyckpath{rer}\!+\!\dyckpath{rre}
    ) \!+\!
        \dyckpath{eee}
    , \dots
    \right\} \nonumber
\end{align}
where, in the last line, we provide a pictorial representation of the colored paths that contribute up to third order. These paths are in one-to-one correspondence with words $w$ satisfying $\varphi(w)=e$ such that the coefficient $[x^L]G_{e}(x)$ in the series is equal to the dimension of the Krylov sector associated to $e$ in a system of size $L$ sites. The asymptotic growth of these coefficients with $L$ is determined by the distance to the nearest singularity~\cite{Flajolet_Sedgewick_2009}. It follows that the size of the largest sector grows as $(1+2\sqrt{m-1})^L$ for large $L$, in contrast to the $(2\sqrt{m-1})^L$ growth of the model that does not include $\tte$ onsite.

\subsection{2D subsystem pair-flip model}

In the 2D subsystem version of the conventional pair-flip model, the only allowed range-2 dynamics are [analogous to Eq.~\eqref{eq:pairflip_free_reduction} in the main text]
\begin{equation}
\begin{bmatrix}
{\color{agen}\ttr} & {\color{agen}\ttr} \\ {\color{agen}\ttr} & {\color{agen}\ttr} \\
\end{bmatrix} 
\longleftrightarrow 
\begin{bmatrix}
{\color{bgen}\ttb} & {\color{bgen}\ttb} \\ {\color{bgen}\ttb} & {\color{bgen}\ttb} \\
\end{bmatrix}
\longleftrightarrow 
\begin{bmatrix}
{\color{cgen}\ttg} & {\color{cgen}\ttg} \\ {\color{cgen}\ttg} & {\color{cgen}\ttg} \\
\end{bmatrix}, 
\label{eqn:sqair-flip}
\end{equation}
because there is no identity character. There are also local moves that do not reduce to a series of range-2 moves. For example, we can ask whether the two configurations
\begin{equation}
\begin{bmatrix}
{\color{bgen}\ttb} & {\color{bgen}\ttb} & {\color{bgen}\ttb} & {\color{bgen}\ttb} & {\color{bgen}\ttb} \\
{\color{bgen}\ttb} & {\color{bgen}\ttb} & {\color{bgen}\ttb} & {\color{bgen}\ttb} & {\color{bgen}\ttb} \\
{\color{bgen}\ttb} & {\color{bgen}\ttb} & {\color{bgen}\ttb} & {\color{bgen}\ttb} & {\color{bgen}\ttb} \\
{\color{bgen}\ttb} & {\color{bgen}\ttb} & {\color{bgen}\ttb} & {\color{bgen}\ttb} & {\color{bgen}\ttb} \\
{\color{bgen}\ttb} & {\color{bgen}\ttb} & {\color{bgen}\ttb} & {\color{bgen}\ttb} & {\color{bgen}\ttb}  
\end{bmatrix} 
\xleftrightarrow{\quad ?\quad}
\begin{bmatrix}
{\color{agen}\ttr} & {\color{bgen}\ttb} & {\color{bgen}\ttb} & {\color{agen}\ttr} & {\color{bgen}\ttb} \\
{\color{bgen}\ttb} & {\color{cgen}\ttg} & {\color{bgen}\ttb} & {\color{bgen}\ttb} & {\color{cgen}\ttg} \\
{\color{bgen}\ttb} & {\color{bgen}\ttb} & {\color{bgen}\ttb} & {\color{bgen}\ttb} & {\color{bgen}\ttb} \\
{\color{agen}\ttr} & {\color{bgen}\ttb} & {\color{bgen}\ttb} & {\color{agen}\ttr} & {\color{bgen}\ttb} \\
{\color{bgen}\ttb} & {\color{cgen}\ttg} & {\color{bgen}\ttb} & {\color{bgen}\ttb} & {\color{cgen}\ttg} 
\end{bmatrix} , 
\end{equation}
are connected by subsystem group dynamics. They have the same subsystem group label (${\color{bgen}\ttb}$ for every row and column) so they are connected by range-4 gates, but cannot be connected by range-2 $\varphi$-preserving gates. While this is analogous to~\eqref{eq:pairflip_linked_loops}, it is larger because of the sublattice structure of the original pair-flip model.
Thus, as in the main text, fully specifying ``subsystem group dynamics'' requires specifying a group $G$ and either an explicit set of $\varphi$-preserving gates or a gate range.

The result is that the subsystem pair-flip model without an $\tte$ shares the phenomenology of the version considered in the main text, i.e., Krylov sector labeling and fractonic dynamics, but with additional complications coming from the sublattice structure.

\section{Further details on dynamical constraints coming from ``flatness''}

In this appendix, we briefly review the construction of robust group word models in 2D~\cite{AlexeyTopologically}. The construction places degrees of freedom on edges of a generic 2D directed lattice, so that edges come with an assigned orientation.  We read words off of paths in the lattice by continuing along a path and reading the character (or its inverse) off of each edge in the path if the edge points along (or against) the path. For example, consider the path that starts at the leftmost vertex of the hexagon,
\begin{equation}
\begin{tikzpicture}[baseline={(0,-0.1)}]
    \begin{scope}[very thick,decoration={markings,
    mark=at position 0.7 with {\arrow{Classical TikZ Rightarrow}}}
    ] 
    \path
    node[regular polygon, regular polygon sides=6, inner sep=14pt] (hex) {};
    \draw [thick, ->, >=stealth] (hex.corner 3) to (hex.corner 2);
    \draw [thick, ->, >=stealth] (hex.corner 2) to (hex.corner 1);
    \draw [thick, ->, >=stealth] (hex.corner 1) to (hex.corner 6);
    \draw [thick, ->, >=stealth] (hex.corner 3) to (hex.corner 4);
    \draw [thick, ->, >=stealth] (hex.corner 4) to (hex.corner 5);
    \draw [thick, ->, >=stealth] (hex.corner 5) to (hex.corner 6);
    \node [circle, agen, fill=agen!20!white, inner sep=0, minimum size=3.75mm, draw, thick, font=\footnotesize] at (hex.side 1) {$\tta$};
    \node [circle, agen, fill=agen!20!white, inner sep=0, minimum size=3.75mm, draw, thick, font=\footnotesize] at (hex.side 3) {$\tta$};
    \node [circle, bgen, fill=bgen!20!white, inner sep=0, minimum size=3.75mm, draw, thick, font=\footnotesize] at (hex.side 4) {$\bar{\ttb}$};
    \node [circle, bgen, fill=bgen!20!white, inner sep=0, minimum size=3.75mm, draw, thick, font=\footnotesize] at (hex.side 5) {$\ttb$};
    \end{scope}
    \end{tikzpicture}
\end{equation}
continues clockwise around, and ends at its beginning. The word on this path is ${\color{agen}\tta} {\color{bgen} \ttb^{-1} \ttb} {\color{agen} \tta^{-1}} = \tte$. Starting at a different point might result in a different element in the same conjugacy class, while reading in the other direction will read the inverse group element, but neither of these have any effect on the identity group element. 

To achieve robust HSF, Ref.~\cite{AlexeyTopologically} first defines \emph{flatness}, the condition that a configuration around a face corresponds to the identity group element. The entire lattice configuration is called flat if all faces are flat.
Then, the word read off of any contractible closed path is also the identity. It is still possible to have nontrivial words on noncontractible loops, and these are the global  labels. The flatness condition requires that two homotopically equivalent paths have the same label, up to conjugacy.

Even if flatness is enforced as a hardcore constraint, there can still be nontrivial dynamics~\cite{AlexeyTopologically}. The minimal constraint-obeying dynamics acts on the edges around a vertex,
\begin{equation}
\label{eq:Av_creates_a_loop}
    \scalebox{0.75}{%
    \begin{tikzpicture}[baseline={(0,0)}]
        \draw [thick, ->, >=stealth] (0:0) to (0:1);
        \draw [thick, <-, >=stealth] (0:0) to (120:1);
        \draw [thick, ->, >=stealth] (0:0) to (240:1);
    \end{tikzpicture}
    }
    \quad
    \xrightarrow{A_v(\tta)}
    \quad
    \scalebox{0.75}{%
    \begin{tikzpicture}[baseline={(0,0)}]
        \begin{scope}[very thick,decoration={markings,
            mark=at position 0.6 with {\arrow[agen,scale=1.25]{Classical TikZ Rightarrow}}}
        ]
        \draw [thick, ->, >=stealth] (0:0) to (0:1);
        \draw [thick, <-, >=stealth] (0:0) to (120:1);
        \draw [thick, ->, >=stealth] (0:0) to (240:1);
        \node [circle, agen, fill=agen!20!white, inner sep=1.05pt, draw, thick, font=\small] at (120:0.51) {$\bar{\tta}$};
        \node [circle, agen, fill=agen!20!white, inner sep=1.5pt, draw, thick, font=\small] at (240:0.51) {$\tta$};
        \node [circle, agen, fill=agen!20!white, inner sep=1.5pt, draw, thick, font=\small] at (0:0.51) {$\tta$};
        \end{scope}
    \end{tikzpicture}
    }
    \, .
\end{equation}
acting as left multiplication by $\color{agen}\tta$ on outward pointing edges and right multiplication by $\color{agen} \tta^{-1}$ on inward pointing edges (see~\cite{AlexeyTopologically} for a detailed definition). Such dynamics are fragmented because they cannot change the labels read off along non-contractible paths, meaning that these global labels label Krylov sectors. In fact, Ref.~\cite{AlexeyTopologically} shows that the HSF is \emph{robust}, in the sense that any constraint-obeying dynamics that act on fewer than $L$ spins at a time preserve the global Krylov labels. 

We can also enforce flatness energetically instead of explicitly. On a given face $f$, the operator
\begin{equation}
    B_f = \sum_{ h_{e_1} \cdots h_{e_k} = \tte} \: \prod_{e \in f} T^{h_e}(e, f) \,
\end{equation}
projects onto flat configurations,
where $T^{h}(e, f)$ evaluates to $\ket{h}\bra{h} $ (or $\ket{h^{-1}}\bra{h^{-1}} $) if the edge $e$ is oriented with (or against) the path around $f$. To enforce flatness, define the diagonal Hamiltonian
\begin{equation}
    H_{{\text{diag}}} = -\Delta \sum_f B_f .
\end{equation}
Ground states are required to be flat, but may have different global labels. Any violation of flatness costs energy $\Delta$, so dynamics below energy $\Delta$ must preserve flatness. Even if the Hamiltonian is subject to generic perturbations, dynamics that change the Krylov label only enter at order $L$ or higher in perturbation theory.  Reference~\cite{AlexeyTopologically} then shows that the fragmentation persists up to a timescale that is exponentially long in inverse perturbation strength. This is the second sense in which the HSF is robust.

We can construct the quantum Hamiltonian
\begin{equation}
    H = -\Delta \sum_f B_f - J\sum_{v, s} A_v(s),
\end{equation}
which includes the off-diagonal $A$ operators. If $G$ is a  finite group and the onsite Hilbert space $\mathcal{A}$ includes every element of the group, then this is Kitaev's quantum double construction~\cite{Kitaev2003}. If $G=\mathbb{Z}$, then this is a $\textrm{U}(1)$ spin liquid, which is unstable to instanton proliferation. For infinite groups other than $\mathrm{U}(1)$, this construction results in Hamiltonians with interesting entanglement~\cite{balasubramanian2023Entanglement, Zhang2024Bicolor} and ground state~\cite{robuster} properties, whose stability requires further study. 

In Kitaev's quantum double and in U(1) gauge theories, flatness leads to an emergent 1-form symmetry. In general, Appendix C in Ref.~\cite{AlexeyTopologically} shows that the resulting higher-form symmetry depends on $G^{\textrm{ab}} = G/[G,G]$, the abelianization of $G$. See Ref.~\cite{AlexeyTopologically} for further details.

\bibliography{refs}

\end{document}